\documentclass[%
 reprint,
 groupedaddress,
 amsmath,amssymb,
 prl,
floatfix,
]{revtex4-2}

\usepackage{comment}
\usepackage{xcolor}
\usepackage{graphicx}
\usepackage{dcolumn}
\usepackage{bm}
\usepackage{physics}
\usepackage{hyperref}

\usepackage{changepage}

\newenvironment{affil}{\begin{adjustwidth}{\parindent}{}\itshape}{\end{adjustwidth}}

\begin{document}

\title{Comment on ``Can Neural Quantum States Learn Volume-Law Ground States?''}

\author{Zakari Denis}
\affiliation{
Institute of Physics, École Polytechnique Fédérale de Lausanne (EPFL), CH-1015 Lausanne, Switzerland
}
\affiliation{
Center for Quantum Science and Engineering, \'{E}cole Polytechnique F\'{e}d\'{e}rale de Lausanne (EPFL), CH-1015 Lausanne, Switzerland
}

\author{Alessandro Sinibaldi}%
\affiliation{
Institute of Physics, École Polytechnique Fédérale de Lausanne (EPFL), CH-1015 Lausanne, Switzerland
}
\affiliation{
Center for Quantum Science and Engineering, \'{E}cole Polytechnique F\'{e}d\'{e}rale de Lausanne (EPFL), CH-1015 Lausanne, Switzerland
}

\author{Giuseppe Carleo}
\affiliation{
Institute of Physics, École Polytechnique Fédérale de Lausanne (EPFL), CH-1015 Lausanne, Switzerland
}
\affiliation{
Center for Quantum Science and Engineering, \'{E}cole Polytechnique F\'{e}d\'{e}rale de Lausanne (EPFL), CH-1015 Lausanne, Switzerland
}

\noindent{\bfseries Comment on ``Can Neural Quantum States Learn Volume-Law Ground States?''}

\bigskip

Passetti \textit{et al.} \cite{passetti_can_2023} recently assessed the potential of neural quantum states (NQS) \cite{carleo_solving_2017} in learning ground-state wave functions with volume-law entanglement scaling. They focused on NQS using feedforward neural networks, specifically applied to the complex SYK Hamiltonian for fermions~\cite{mon1975}. Their numerical results hint at an exponential increase in the required variational parameters as the system size grows, apparently tied to the entanglement growth within the SYK ground state. This challenges the general utility of NQS for highly entangled wavefunctions, contrasting with established analytical~\cite{deng2017,gao2017,glasser2018,levine2019,sharir2022,sun2022} and numerical findings~\cite{deng2017,sun2022}.
Based on our experiments, we show that suitably chosen NQS \emph{can} learn ground states with volume-law entanglement both for spin and fermionic problems. We argue that the setup utilized in \cite{passetti_can_2023} reveals the inefficiency of non-fermionic NQS to learn the sign structure of fermionic states, rather than a general connection between entanglement content and learnability hardness.

\begin{figure*}[th!]
    \centering
    \includegraphics{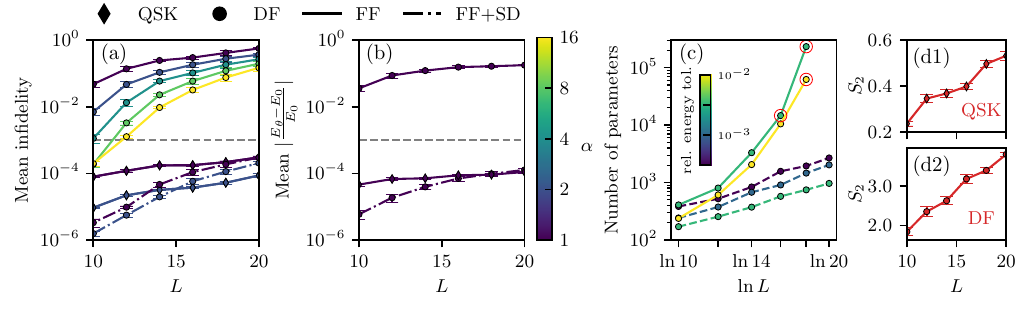}
    \caption{(a) Infidelity as a function of system size for fidelity-optimization simulations on the ground state of the QSK (diamonds) and DF (circles) Hamiltonian with a two-layer feedforward network of varying $\alpha$. (b) Corresponding relative error of the variational energy for $\alpha=1$. (c) Number of variational parameters required to match a given relative error on the ground-state energy as a function of system size. Red circles denote points obtained via extrapolation. (d1-d2) Volume-law scaling of the Rényi-2 entanglement entropy for the maximum bipartition computed on the ground state of the QSK and DF Hamiltonians. Error bars denote the standard error on the mean for at least ten random realizations of the Hamiltonian.}
    \label{fig}
\end{figure*}

In Fig.~\ref{fig}(a), we show the average infidelity obtained by performing fidelity optimization on the ground-state wavefunction of ten instances of the quantum Sherrington-Kirkpatrick model (QSK)~\cite{schindler2022}. %
We show results for a two-layer perceptron NQS with real parameters, $\tanh$ activation, and varying hidden-unit density $\alpha$. Although the ground state of the QSK model exhibits volume-law entanglement~\cite{schindler2022} (see also Fig.~\ref{fig}(d1)), quadratically increasing the number of parameters with system size $L$ proves sufficient to achieve infidelity below a threshold of $10^{-3}$, a more stringent condition of learnability than the energy relative error discussed in Ref.~\cite{passetti_can_2023}, also shown in Fig.~\ref{fig}(b). This already shows that NQS \emph{can} efficiently learn non-trivial, volume-law entangled ground states, thus answering the question posed in the title of Ref.~\cite{passetti_can_2023} in the positive and consistent with previous literature in the field~\cite{deng2017,gao2017,glasser2018,levine2019,sharir2022,sun2022}.

We also consider a SYK-type disordered fermionic Hamiltonian (DF)~\cite{dieplinger2021}: $\hat{H} = \sum_{i<j}(J_{ij} \hat{c}_i^\dagger\hat{c}_j^{\mathstrut} + V_{ij}\hat{n}_i\hat{n}_j)$, with random symmetric matrices $J_{ij}$ and $V_{ij}$ drawn from $\mathcal{N}(0,1/\sqrt{L})$. This Hamiltonian yields ground states with volume-law entanglement (Fig.~\ref{fig}(d2)). In Fig.~\ref{fig}(a), (b) and (c), we compare two feedforward neural-network architectures: (i) a two-layer perceptron similar to Ref.~\cite{passetti_can_2023}, and (ii) an architecture with output dimension $N{\times}L$, used in a Slater determinant through the backflow-transformation prescription \cite{luo2019}. Unlike the latter, the former fails to learn the target ground state within a required infidelity of $10^{-3}$. In Fig.~\ref{fig}(c), we compare the scaling of the number of parameters to match a desired error. For model (i), we reproduce the exponential scaling observed in \cite{passetti_can_2023} for the SYK Hamiltonian. For model (ii), restoring the proper sign structure through the determinant makes the complexity drop from exponential to only polynomial. The scaling behavior observed in (i) also occurs in the non-interacting ($V_{ij}=0$) case (not shown here), whereas fermionic NQS (ii) learn these states efficiently. These findings suggest that the hints of exponential scaling observed in \cite{passetti_can_2023} with a type-(i) Ansatz are unrelated to entanglement content; rather, they are an intrinsic manifestation of the inefficacy of a \emph{non-fermionic} NQS in describing a \emph{fermionic} system.

In summary, our results demonstrate that wisely chosen NQS are effective at learning ground states characterized by volume-law entanglement in strongly interacting spin and fermionic systems. While Ref.~\cite{passetti_can_2023} emphasizes the putative inability of NQS to learn highly entangled ground states efficiently, we have shown here that their findings are instead strictly unrelated to the entanglement content of the states to be learned. Rather, they show that pure feed-forward architectures are not capable of efficiently describing strongly correlated fermionic ground states. We remark, however, that this very observation is what motivated the development of NQS specifically tailored for strongly interacting lattice fermions~\cite{nomura_restricted_2017, luo2019, robledo_moreno_fermionic_2022} over the past six years.  

\begin{acknowledgments}
This work was supported by SEFRI under Grant No.\ MB22.00051 (NEQS - Neural Quantum Simulation). NQS simulations were performed using NetKet~\cite{netket3:2022}, the entanglement entropy was evaluated using QuTiP~\cite{johansson2013}. 
The code for the simulations can be found in \cite{repo}. 
\end{acknowledgments}

\medskip

\noindent{}Z. Denis, A. Sinibaldi, and G. Carleo

\begin{affil}
Institute of Physics, École Polytechnique Fédérale de Lausanne (EPFL), CH-1015 Lausanne, Switzerland
\end{affil}

\begin{affil}
Center for Quantum Science and Engineering, \'{E}cole Polytechnique F\'{e}d\'{e}rale de Lausanne (EPFL), CH-1015 Lausanne, Switzerland
\end{affil}

\bibliography{references}%

\end{document}